\documentclass[10pt]{article}
\usepackage{amssymb}
\usepackage{amsmath}
\usepackage{color}
\usepackage{graphicx}

\def\be{\begin{equation}}
\def\ee{\end{equation}}
\def\bea{\begin{eqnarray}}
\def\eea{\end{eqnarray}}

\def\<{\langle}
\def\>{\rangle}
\def\~{\tilde}

\DeclareGraphicsExtensions{.png}

\numberwithin{equation}{section}
\numberwithin{figure}{section}
\numberwithin{theorem}{section}
\numberwithin{teorema}{section}

\def\be{\begin{equation}}
\def\ee{\end{equation}}
\def\bc{\begin{center}}
\def\ec{\end{center}}

\title{\bf Some thoughts on the ontogenesis in B-cell immune networks}
\author{Adriano Barra$^1$,  Silvio Franz$^2$, Thiago  Sabetta$^3$}
\begin{document}
\date{}
\maketitle

\begin{center}
{\small
\vskip-0.5cm

%Address of the first author
\footnote{e-mail:{\tt adriano.barra@roma1.infn.it}} Dipartimento
di Fisica, Sapienza Universit\`a di Roma \vskip-0.5cm
\footnote{e-mail:{\tt silvio.franz@itcp.trieste.it}} Laboratoire
de Physique Th$\acute{e}$orique et Mod$\grave{e}$les Statistiques,
Universit$\acute{e}$ Paris-Sud $11$ \vskip-0.5cm
\footnote{e-mail:{\tt thiago.penteado-sabetta@polytechnique.edu }}
$\acute{E}$cole Polytechnique, Paris}

\end{center}

%\vskip 0.5cm

{\small \bf------------------------------------------------------------------------------------------------------------}

\vskip-1cm
{\small

{\bf Abstract.}
%{\bf Sunto.}
%\vskip-0.5cm
%
%{\it Title in Italian.}  Here the abstract in Italian ...
%
}

%We use a statistical mechanics approach to investigate the effect
%of "ontogenesis" on the structure of B-cell idiotypic immune
%network. Within a simple scheme for B-cell behavior, we show that
%early interactions of immature cells with self-antigens can have a
%strong influence on the connectivity distribution of the mature
%network. With respect to random network (previously investigated),
%we find an increase both in the average connectivity and its
%variance.

%Poi avevi cancellato a pag. 2 la seconda riga: and the plethora...
%...countless e sostituito "eaten" con destroyed. C'è anche
%un'aggiunta lla formula (3.3) che però non capisco bene; tipo:
%\frac{e^{\phi\sigma}}{2cosh\phi}.

We are interested in modeling theoretical immunology within a
statistical mechanics flavor: focusing on the antigen-independent
maturation process of B-cells, in this paper
 we try to revise the problem of  self vs non-self discrimination
by mature B lymphocytes. We consider only B lymphocytes: despite
this is of course an oversimplification, however such a toy model
may help to highlight features of their interactions otherwise
shadowed by main driven mechanisms due to i.e. helper T-cell
signalling.
\newline
By analyzing possible influences of the ontogenesis of the immune
system on the final behavior of B lymphocytes, we try to merge
over the purely negative selection mechanism at their birth with
the adult self-regulation process. The final goal is a
"thermodynamical picture" by which both the scenarios can exist
and, actually, be synergically complementary: Trough numerical
simulations we impose on a recent scheme for B-cell interactions,
that part of self-reactive lymphocytes are killed during the
ontogenesis by which two observations stem: At first the so built
system is able to show anergy with respect to the previously
encountered self even in its mature  life, then this naturally
leads to an increasing variance (and average) in the connectivity
distribution of  the resulting idiotypic network. As a
consequence, following Varela perspective, this shift may
contribute to push to anergy those self-directed cells which are
free to explore the body: identifying the latter as the highly
connected ones, anergy is imposed even via the B-network
regulation, and its strength is influenced by the negative
selection.

%\vskip-0.6cm

%{\small
%\bf------------------------------------------------------------------------------------------------------------}
\cleardoublepage

%\tableofcontents

\section{Introduction}

Immunology is probably one of the fields of science  which is
experiencing the greatest amount of discoveries in these decades:
As the amount of works increases, the need for minimal models able
to offer a general, coarse grained, framework where these may find
a collocation is a must for modelers interested in this field.
\newline
Despite actors in the immune system are many, for the sake of
simplicity, we are going to focus only on the B-cell ensemble and,
for the sake of clearness, we allow ourselves in presenting a
streamlined introduction to the main concepts on their world and a
state of the art in self/non-self discrimination, on which we will
rely soon.
\newline
The purpose of the immune system is to detect and neutralize the
molecules, or cells, dangerous for the body (antigens, which could
be foreign invaders - e.g. viruses or bacteria - or deranged  -
e.g. cancerous - cells of the host), without damaging healthy
cells \cite{a11}. The humoral response performed by B lymphocytes
consists in analyzing the antigen by each family of identical
B-cells (clones), then the one/s with the best matching
antibody\footnote{The first postulate of immunology states that,
hypersomatic mutation apart \cite{a25,a26}, a given clone of
B-cells -namely a family of identical lymphocytes- produces always
the same antibody.} undergoes clonal expansion and releases its
immunoglobulins (clonal selection theory): the latter are able to
bind the pathogens and neutralize their chemical abilities; then,
the resulting complex is destroyed  by macrophages and order
established again.
\newline
For achieving this goal, the immune system needs an enormous
number of different clones, each one having a particular receptor
for antigens. As these receptors are generated randomly by somatic
mutation at the genetic level, the body may produce both
antibodies attaching to (a part of the) intruders (i.e. viruses),
as well as to internal ones (self reactive lymphocytes), which, if
not carefully checked, may induce autoimmunity, an obviously
unwanted feature.
\newline
To avoid this failure, at least two mechanisms are thought to work
(for self/non-self discrimination), at different levels in the
immune systems as we are going to resume.
\newline
B-cells are generated, and maturate, in the bone marrow, where
they are exposed to "negative selection rule"\footnote{We only
stress here strong differences among B-cell maturation in the bone
marrow and T-cell one in the thymus, due to the lacking of TCR by
the humoral effectors \cite{kuby}\cite{perelson1}\cite{perelson2}.
Unlike  TCR that evolved to recognize characteristic patterns of
pathogens, BCR on B-cells is primarily diversified in random
fashion and has not evolved to recognize a particular structure.
Therefore each B cell can not discriminate self versus non self
alone \cite{kitamura}.}: In a nutshell, driven by the nurse-cells,
these lymphocytes are made to interact with an (available)
repertoire of self-antigens, namely molecules/cells belonging to
the host body, and those who are found to respond to them (so
potential autoimmune B-cells) experience induced apoptosis, such
that only B-cells unable to attach to the available self survive
and share the freedom of exploring the body thereafter \cite{a11}.
\newline
It is in fact widely accepted that the bone marrow produces daily
$\sim 10^7$ B cells, but only $\sim 10^6$ are allowed to circulate
trough the body, the remaining $90\%$ undergoing apoptosis because targeted as self-reactive \cite{science}:
as shown for instance by Nemazee and Burki \cite{NB}, this depletion
of the potential defense is due to the negative selection (clonal
deletion) of immature B-cells expressing self-reactive antibodies.
\newline
However as only a fraction of self-antigens are present into the
bone marrow, self-reactive lymphocytes not expressing specific
receptors (BCR)  against the
available self are allowed to circulate freely by this first
security procedure: another mechanism must act at peripherals
levels (i.e. in the lymphonodes, spleen and lever).
\newline
Coherently, after their experiments, Goodnow  was been able to
show \cite{good} that these self-reactive lymphocytes indeed exist
in the body, but instead of undergoing apoptosis, they experience
anergy in their responses (namely, under the proper stimulus, they
do not responde). Furthermore his experiments showed that this
anergy could be related to the corresponding low expression of IgM
on the external membrane of the self-reactive lymphocytes
(establishing a ratio $1:20$ with an ordinary one) implicitly
suggesting both an effect in their response function by the
network\footnote{The term "network" here is  meant to include the
whole immune system interactions, not just the B-core, so at first
exchanges among B-cells and T-cells, cytokine messengers and so on
\cite{a11}.}, as well as  that a reduction in the expression level
of surface receptors IgM quantitatively resulted via a biased
signal transduction producing the anergic state.
\newline
Furthermore, still highlighting the need for a second pathway of
control, experimentally, in a healthy body, a low dose of (some
of) self-directed antibodies is commonly found (negative selection
is not exhaustive), and theoretically (highlighting the importance
of the B-cell network alone), if the amount of information needed
to tackle a response (i.e. the amounts of epitopes) is believed to
range in the order $\sim 10^3$\cite{perelson3}, within a pure
action-reaction
 approach, the immune system would need $O(2^{10^{3}})$ different clones,
 which is an enormous number with respect to the amount of actual
 ones found in the body (i.e. $\sim O(10^{10})$ \cite{perelson3}): information
for pattern recognition must be spread over a network of
interacting B-cells.
\newline
So, self reacting  clones that have not been eliminated in the
bone marrow become unresponsive to (self)-antigen, which is termed
"anergy": In  these cells, continued binding of self-antigen is
required to be kept in anergic state \cite{gauld}\cite{goodnow2}.
\newline
Following Kitamura \cite{kitamura},  a key in understanding
 the strategy by which B-cells manage self/non-self
discrimination (despite the very incomplete knowledge of the BCR signaling
pathways)  is their double signalling activation need (namely
the presence of the antigen and the stimulation by the cognate T
helper): while the double signaling induces clonal expansion, only
one signal (the [self]-antigen) may induce a suppression. As a
consequence, B-cell network may synergically use the helper
T-cells both for activation against pathogens as well as for
anergy induction with respect to self ones (at T-helpers are equipped by TCR).
\newline
Furthermore, the anergic  B cell shows several features that
characterize its "peculiarity": cell surface expression level of
BCR (IgM) is reduced and that of $CD5$ (IgD) increased, lifespan
is shortened and entry into the lymphoid follicles is
prohibited\footnote{This suggests that anergic B cells, in
contrast to non-self reactive B cells, fail to compete for
survival of chemotactic factor \cite{cyster}.}. Their BCR are
desensitized and therefore the B cell do not proliferate in
response to antigens even in the (not usual) presence of cognate
T-cell help (which implicitly may suggest other mechanisms then
the need of double signaling by antigen and helper alone) but are
instead made anergic or eliminated by Fas-induced apoptosis
(AICD).
\newline
Despite this may actually be the main strand  for explaining
self/non-self discrimination, other mechanisms may cooperate, and,
among these, as experiments in vitro with the Jerne network are
prohibitive by construction, we plan to investigate them trough
statistical mechanics simulations: the goal is to isolate this
possible path from the main one and see if it can contribute to
the overall regulation.
\newline
As a consequence we spend a few words on this network:
\newline
The idea of this internal B-core network appeared early in immunology
\cite{a8}, and its concretization happened when Jerne \cite{a44},
in the $70$'s, suggested that each antibody must have several
idiotopes which are detected by other antibodies. Via this
mechanism, an effective structure of interacting antibodies is
formed, in which the latter not only detect antigens, but also
function as individual internal images of them and are themselves
detected and acted upon. These network interactions provide a
"dynamical memory" for the immune system, by keeping the
concentrations of antibodies (especially those representing
encountered antigens) at appropriate levels. This can be
understood as follows: At a given time a virus is introduced in
the body and starts replication. At high enough concentration, it
is found by the proper B-lymphocyte counterpart: let us consider,
for simplicity, a virus as a string of information (i.e.
$1001001$) \footnote{The dichotomy of a binary alphabet in strings
mirrors the one of the electromagnetic field governing chemical
bonds.}. The complementary B-cell producing the antibody Ig1,
which can be thought of as the string $0110110$ then will start a
clonal expansion and will release high levels of Ig1. As a
consequence, after a while, another B-cell will meet $0110110$
and, as this string never (macroscopically) existed before,
attacks it by releasing the complementary string $1001001$, that,
actually, is a "copy" (internal image) of the original virus but
with no DNA or RNA charge inside \footnote{This counter-images
have been revealed experimentally in several researches, i.e.
\cite{cazenave}.}. The interplay among these helps in keeping
memory of the past infections.
\newline
Once a network theory is achieved, it is easy to understand that,
given the "hyper-fine" recognition mechanism, this implies the
connectivity of such a network to range over several orders of
magnitude: as a result a biological interpretation is handily and
originally due to Varela and Coutinho \cite{a30,a38,a39}: nodes
(i.e. clones) which are poorly connected are thought of as
antigen-directed as they can easy respond to external fields
(roughly speaking are more approximable as single particles),
while nodes that are highly connected can probably be
self-directed as they can be strongly influenced by the (large
amount of) nearest neighbors, which, in this case, may keep them
in a state of anergy.
\newline

\section{The minimal model}

Focusing on these "emergent properties" of B-cell networks, (and
neglecting investigation during B-cell's birth trough the negative
selection), inspired by pioneering ideas of thermodynamical flavor
\cite{giorgio} in a recent series of papers \cite{AB1,AB2,AB3} a
statistical mechanics model for such systems has been introduced.
Within that framework the distinction among self and non-self was
thought of at a cooperative level alone, in a pure Varela style:
no ontogenesis were investigated and no learning rules discussed,
while it was shown how to obtain a scale free weighted
connectivity distribution from a wide class of antibody's
interactions, as a benchmark for self-non self discrimination a
posteriori.
\newline
Despite a scenario able to recover several real features of the
immune system was  already achieved in this way, however, a
complete elimination of a learning process during the ontogenesis
was
 unrealistic \cite{kuby} and indeed its existence could alter the mature network functionalities:
so we want to move over and show that the two pictures discussed
in the introduction (negative selection and idiotypic network
regulation) may act synergically and naturally accounted by the
model itself.
\newline
Taking an antibody as a binary vector made up of the possible
expression of $L$ idiotopes, we assume that these can be thought
of as strings of the same length\footnote{The molecular weight for
each Igs is accurately close to $15\cdot10^4$ and each idiotope on
average is large as each other (see \cite{nobel})}, such that, as the
elementary $L$ idiotopes can be introduced as,
\begin{equation}  \xi_1 = (1,0,0,...,0), \
 \xi_2 = (0,1,0,...,0),  \ ..., \  \xi_L = (0,0,0,...,1),
\end{equation}
forming an orthogonal base in the $L$-dimensional space $\Upsilon$
of the antibodies,  in such a way that the generic $i^{th}$
antibody $\xi^i$ can then be written as a linear combination of
these eigenvectors $\{\xi^{i} \} = \lambda_1^{i} \xi_1,
\lambda_2^{i} \xi_2, ..., \lambda_L^{i} \xi_L$, with
$\lambda_{\mu}^i \in (0,1)$ accounting for the expression $(1)$ of
a particular $\mu$th idiotope or its lacking $(0)$.
\newline
In this way, as often done in modern modeling of antibody
affinities \cite{a5}\cite{a67}, we relax the earlier simplifying
assumption of "a perfect mirror of a mirror" for the interacting
immunoglobulins simply asking that the better the matches among
idiotopes, the stronger the stimulus occurring between the
respective clones via these messengers.

Moreover, the system is made up of an ensemble of $N$ different
clones, each composed of $M$ identical lymphocytes; a given
lymphocyte $i$ (whose corresponding antibody is $\xi_i$), is then
described by the dichotomic variable $\sigma_i^{\alpha} = \pm 1$,
with $\alpha=1,...,M$, and $i=1,...,N$, such that the value $-1$
denotes an anergic/absent state (low level of antibodies
secretion) while the value $+1$ a firing state (high level of
antibodies secretion).
\newline
To check immune responses we need to introduce the $N$ order
parameters $m_i$ as local magnetizations:
\be m_i(t)= \frac{1}{M}\sum_{\alpha=1}^{M}\sigma_i^{\alpha}(t).
\ee
From the magnetizations $m_i \in [-1,1]$, which play the role of
the principal order parameters, we can define the concentrations
of the firing lymphocytes belonging to the $i^{th}$ family as
\footnote{The bridge among concentrations in chemical kinetics and
statistical mechanics has been early investigated by Thompson
\cite{thompson} in the context of red cells, but the same should
hold even for the white ones and, however, does not affect our
investigation.}:
\begin{equation} \label{eq:concetration}
c_i(t) \equiv \exp\left[ \tau \frac{(m_i(t)+1 )}{2} \right], \ \
\tau = \log M.
\end{equation}
Further we introduce the Hamiltonian $H$ which encodes the
interactions among lymphocytes as well as the interactions among
lymphocytes and the external antigens:
\begin{equation}\label{eq:total_H}
H=H_1 +H_2  =  -N^{-1} \sum_{i<j}^{N,N} J_{ij} m_i m_j - c
\sum_k^N h_k m_k,
\end{equation}
where $c$ rules the amount of the external antigen present in the
host and $h$ its epitopal characteristics, whose links with the
antibodies will be discussed hereafter:
\newline
In fact, we must briefly resume how the interaction matrix
$J_{ij}$ is built up (and consequently $h_k$): Given two strings
$\xi_i$ and $\xi_j$, their $\mu$-th entries are said to be
complementary, iff $\xi_i^{\mu} \neq \xi_j^{\mu}$. As each entry
$\mu$ of the $i$-th string (Ig) is extracted randomly according to
the discrete uniform distribution in such a way that
$\xi_i^{\mu}=1$ ($\xi_i^{\mu}=0$) with probability $1/2$, given a
couple of clones, say $i$ and $j$, therefore the number of
complementary entries $c_{ij} \in [0, L]$ can be written as \be
c_{ij} = \sum_{\mu = 1}^{L} [\xi_i^{\mu} (1 - \xi_j^{\mu}) +
\xi_j^{\mu} (1 - \xi_i^{\mu}) ]. \ee

The affinity between two antibodies is expected to depend on how
much complementary their structures are. In fact, the non-covalent
forces acting among antibodies depend on the geometry, on the
charge distribution and on hydrophilic-hydrophobic effects which
give rise to an attractive (repulsive) interaction for any
complementary (non-complementary) match. Consequently, we assume
that each complementary / non-complementary entry yields an
attractive / repulsive contribute. In  general, attractive and
repulsive contributes can have different intensity and we quantify
their ratio with a parameter $\alpha \in \mathbb{R}^+$. Hence, we
introduce the functional $f_{\alpha,L}: \Upsilon \times \Upsilon
\rightarrow \mathbb{R}$ as \be \label{eq:affinity}
f_{\alpha,L}(\xi_i,\xi_j) \equiv
 [\alpha c_{ij} - (L-c_{ij})], \ee
 which provides a simple measure of how ``affine'' $\xi_i$ and $\xi_j$ are.
In principle, $f_{\alpha,L}(\xi_i,\xi_j)$ can range from $-L$
(when $\xi_i = \xi_j$) to $\alpha L$ (when all entries are
complementary, i.e. $\xi_i = \bar{\xi}_j$). Now, when the
repulsive contribute prevails, that is $f_{\alpha,L} < 0$, the two
antibodies do not match each other and the coupling among the
corresponding lymphocytes  $J_{ij} (\alpha,L)$ is set equal to
zero, conversely, we take $J_{ij} (\alpha,L)=
\exp[f_{\alpha,L}(\xi_i,\xi_j)] / \langle \tilde{J}
\rangle_{\alpha,L}$, being $\langle \tilde{J} \rangle_{\alpha,L}$
the proper normalizing factor so to keep unitary the average
coupling.
\newline
Hence, nodes can interact pairwise according to a coupling
$J_{ij}(\alpha,L)$, which is defined as:
\begin{equation} \label{eq:J} J_{ij} (\alpha,L)
\equiv \Theta(f_{\alpha,L}(\xi_i,\xi_j)) \frac{\exp
[f_{\alpha,L}(\xi_i,\xi_j)] }{ \langle \tilde{J}
\rangle_{\alpha,L} },
\end{equation}
where $\Theta(x)$ is the Heaviside function returning $1$ if
$x>0$, and $0$ if $x \leq 0$; notice that the affinity matrix is
symmetric, namely $J_{ij}(\alpha,L)=J_{ji}(\alpha,L)$ \footnote{We
stress that, built in this way the affinity matrix, we can
construct a dynamics respecting detailed balance \cite{thompson},
as a result relaxation to Maxwell-Boltzmann distribution in
ensured an we can use standard MonteCarlo techniques in
simulations.}. The coupling $h_i^k$ between the antigen
$\bar{\xi}_i$ and the antibody $\xi_k$ is defined analogously.
\newline
From a statistical mechanics perspective, the Hamiltonian is the
average of the "energy" inside the system and thermodynamic
prescription is that the system tries to minimize it. As a
consequence, according to $H_1$, increasing the antigen
concentration makes the antibody response grow such that if
$c(t_2)> c(t_1)$, with $t_2>t_1$, the same happens for each
involved clone $m(t_2)>m(t_1)$ and viceversa. Moreover, according
to $H_2$, two generic clones $i$ and $j$ in mutual interactions,
assuming here $J_{ij}
> 0$, tend to imitate one another, i.e. if $i$ is quiescent, it
tries to make $j$ quiescent as well (suppression), while if the
former is firing it tries to make firing even the latter
(stimulation), and symmetrically $j$ acts on $i$.
\newline
At this stage we deal with a mature immune system whose order
parameters, namely the magnetizations, are centered symmetrically
distributed. Despite agreement with phenomenology
\cite{AB1}\cite{AB2}\cite{AB3}, thinking that the absence of a
stimulus can be understood as a stimulus, seems difficult to be
explained and surely not exhausted here \footnote{We remember
however that "similar" mechanisms initially difficult to be
understood have already appeared several times in the scientific
literature, i.e. the paradigmatic  negative solutions of the Dirac
equation led to the concept of lacunaes in quantum mechanics.}.
However if we write down the internal field acting on the generic
$i^{th}$ clone, calling it $\varphi_i$ and labeling the weighted
connectivity of such a node as $w_i = \sum_j J_{ij}$, we see that
$$
\varphi_i = \sum_j J_{ij}m_j + h_i,
$$
where we meant the antigen with $h_i$ (and set it to zero now for
clearness, namely $h_i=0$). If we now switch to the concentrations
(through eq. $2.3$), we see that this field can be written as
$$
\varphi_i = \frac{2}{\tau}\sum_j J_{ij} \log c_j - w_i.
$$
From a B-cell concentration viewpoint each clone experiences in
this way a contribution from the weighted connectivity of the
idiotypic network that pushes to anergy, furthermore, the larger
the connectivity of the node, the stronger the resulting
imposition to anergy\footnote{Low connectivity inhibition
experienced by the non-self directed clones accounting for the low
dose  tolerance phenomenon \cite{a11}\cite{giorgio}.}.

\section{Ontogenesis}

As we mentioned, during the ontogenesis of the repertoire of
B-cells, those interacting with self-antigens undergo to negative
selection (roughly speaking are killed\footnote{The  biological
motivation of this initialization is that B-cell antigen receptor
signal transduction machinery transiently activate the cell but
rapidly endocytosis any antigen that bind BCR to ensure the
cessation of the initial activation of the cell. Immature B-cells
(during ontogenesis) are not yet equipped for the antigen
presentation, furthermore, no helper T-cells are available to
double signaling the activation process, the whole pushing the
cell to induced apoptosis.}). As a consequence, here we
implemented the following learning rule: At the beginning, and
once for all, $N_S$ vectors coding for their corresponding
antigens (randomly drawn from a uniform distribution) are
arbitrarily labeled as "self" and stored into the algorithm.
\newline
Then, each creation time\footnote{The genesis of B-cells should be
a continuous time process in the bone marrow, however, dealing
with numerical simulations, we discretize the time such that each
time iteration a -fixed- amount of lymphocytes is generated.} a
set of $P<N$ newborn lymphocytes is generated and each of these
$P$ B-cells is made to interact with all the $N_S$ self-antigens:
those who are able to bind this available self (namely, display a
positive affinity) are killed.
\newline
As a consequence only a fraction $P^*<P$ is
retained. Then another set of $P<N$ of newborn lymphocytes is
randomly created and the whole ensemble of $P+P^*$ of lymphocytes
 is made to interact with the $N_S$ self-antigens. Again those able
to bind the self are eliminated. The process stops when a
size-desirable ensemble of lymphocytes is created (i.e. a
repertoire of size $N$).
\newline
Once this ontogenetic process finishes, we study the property of
the network obtained in this way and compare them with respect to
a network resulting in a purely random fashion without any
learning rule.

\subsection{Numerical implementation}

To test the features of a so generate artificial immune network, we use Monte
Carlo simulation: Following a standard statistical mechanics
approach \cite{a94} the dynamics can be written as
\begin{equation}\label{markov}
\sigma_i^{\alpha}(t+1) = sign \left( \tanh(\beta \varphi_i(t)) +
\eta_i^{\alpha}(t)\right ),
\end{equation}
where $\varphi_i(t)$ is the overall stimulus felt by the $\alpha$
lymphocyte of the $i$ clone , namely
\begin{equation}
\varphi_i(t) = N^{-1} \sum_j^N J_{ij} m_j(t) + h_i(t),
\end{equation}
and the randomness is in the noise implemented via the random
numbers $\eta_i^{\alpha}$, uniformly drawn over the set $[-1,+1]$.
 The impact of this noise on the state
$\sigma_i^{\alpha}(t+1)$ is tuned by $\beta$, such that for
$\beta=\infty$ the process is completely deterministic while for
$\beta=0$ it is completely random.
\newline
As the affinity matrix is symmetric, for this detailed balanced
system, the sequential stochastic process (\ref{markov}) can be
implemented on a machine via Glauber dynamics, with the following
expression for the transition rate $W_i$  \be
W_i(\sigma_i^{\alpha}) = \Big( 1 + \exp(\beta \Delta
H(\sigma_i^{\alpha}; \xi)) \Big)^{-1}, \ee where $\Delta
H(\sigma_i^{\alpha}; \xi) = H(F_i^{\alpha} \sigma_i^{\alpha}; \xi)
- H(\sigma_i^{\alpha}; \xi)$ and $F_i^{\alpha}$ is the "spin-flip"
operator that reverses $\sigma_i^{\alpha} \to -\sigma_i^{\alpha}$.
\newline
Using these probability rates, it is immediate to define a Monte
Carlo scheme (MC) for simulating the network: The general system
setup is made of by $N, N_S, N_A$ elements, where $N$ is the
amount of the (mature) repertoire, $N_S$ the amount of learned
self-antigens and $N_A$ the amount of available external antigens.
We study the response of the system against  the global
field composed always by $N_S$ self antigens and a variable amount
from the $N_A$ ensemble. The chosen dynamics is the standard
Metropolis where each MC iteration is built by $N \cdot M$ steps
(i.e. the amount of cellular automata in the system): for each of
these steps one lymphocyte $\sigma_i^{\alpha}$, randomly drawn
over the repertoire, is chosen and flipped $\sigma_i^{\alpha}
\rightarrow -\sigma_i^{\alpha}$: the variation into the energy
term $\Delta H(\sigma;\xi)$ is then evaluated and if this delta is
negative such a trial move is retained by the system, otherwise
randomly rejected with probability $\propto \exp(-\beta
H(\sigma;\xi))$.
\newline
The value of $\alpha$ is kept to $\alpha=0.7$ following biological
matching as explained in \cite{AB1}.

 \subsection{Results: Self-tolerance, memory and saturation.}

At first we stress that, as simulations with a realistic amount of
clones are still too heavy in CPU time consuming,  we worked at
various repertoire sizes but we tested the robustness of the
results trough a finite size scaling which is reported in
Fig.$3.1$ (right).
\newline
Once the repertoire has been created (and an established network
of interacting B-cells achieved), external antigens are presented
to it and responses are checked. At first, since we constructed
the repertoire with the intent of tolerance to self-antigens, we
check its robustness by presenting to the system a field composed
only by self-antigens: at low temperature, anergy to self is
completely fulfilled (not shown in plots), for each experienced
field made of by $1,..,N_S$ self-antigens.
%\begin{figure}[tb]
%\resizebox{1.00\columnwidth}{!}{\includegraphics{sette.eps}}
%%\resizebox{0.45\columnwidth}{!}{\includegraphics{IsteMatheB15B08A01.eps}}
%\caption{blablabla.} \label{sette}
%\end{figure}%
\begin{figure}[tb]
\resizebox{0.50\columnwidth}{!}{\includegraphics[angle=-90]{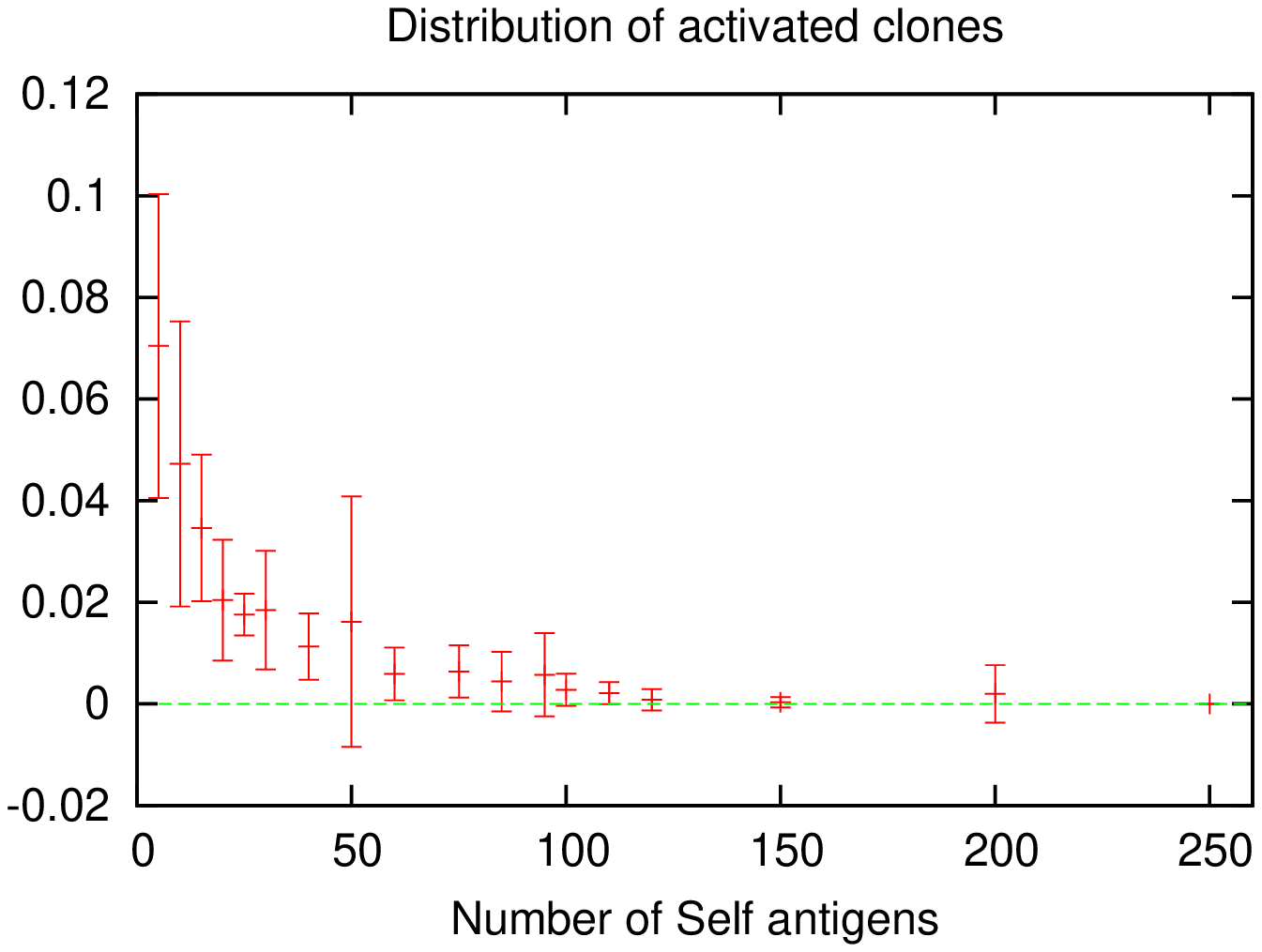}}
\resizebox{0.50\columnwidth}{!}{\includegraphics[angle=-90]{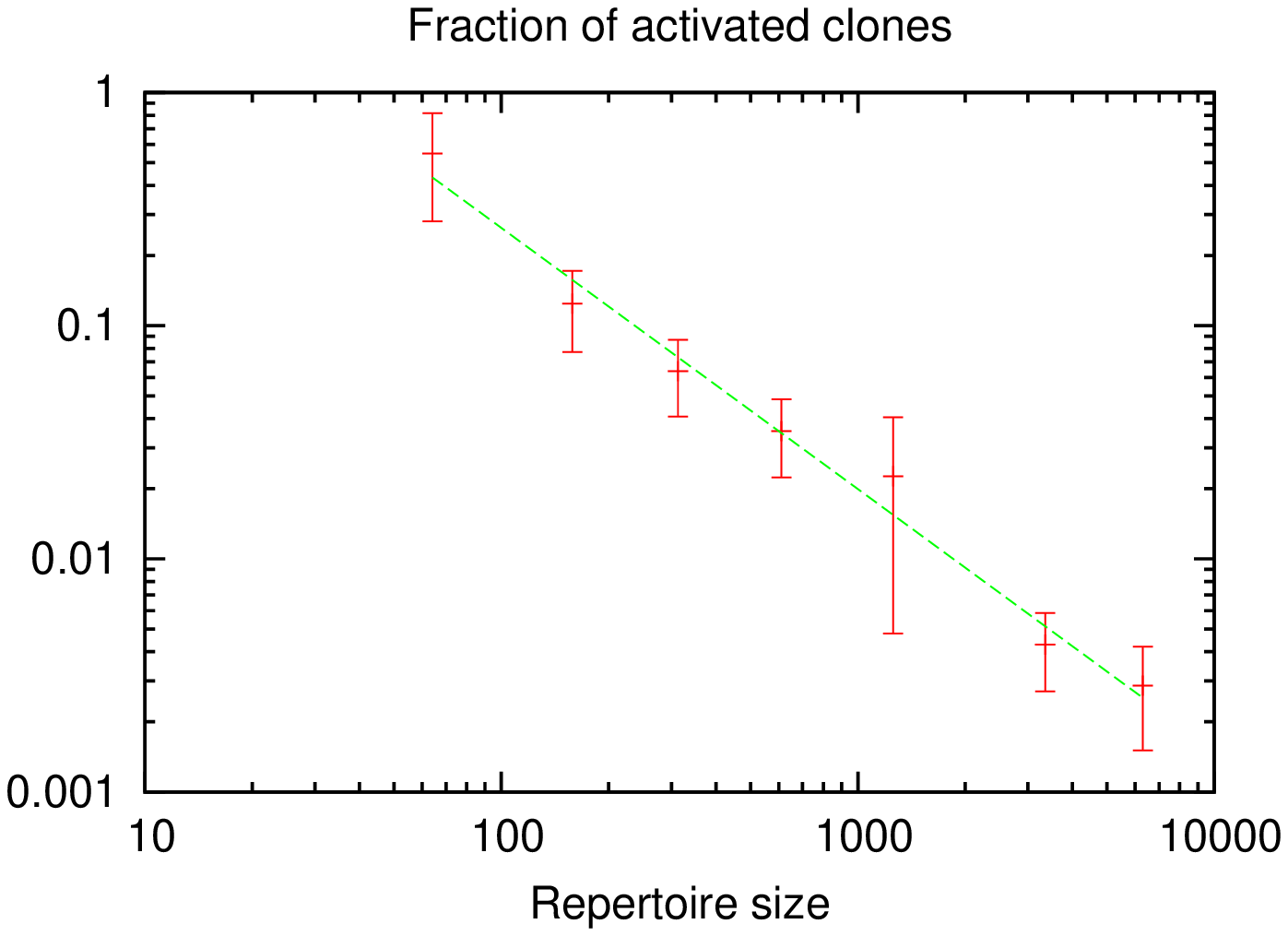}}
\caption{Left: Distribution of the activated clones for an immune
network at rest built up by $N=628$ clones versus the amount of
self-antigens used to generate the repertoire with antibodies made
of by strings of $L=11$ epitopes. Right: Finite size scaling of
the system. Averaged response of the network created trough a
repertoire with $L=8,...,14$ epitopes (keeping the fractions of
the present clones and self-antigens constant) against one
(randomly chosen) antigen of the repertoire itself. Coherently
with the request that only a finite fraction of clones remains
active increasing the network size, the fit is obtained trough
$O(N^{-1})$ power (the exact value of the fit with $N^x$ gives $x
\sim -1.12$).} \label{scaling}
\end{figure}
\begin{figure}[tb]
\resizebox{0.50\columnwidth}{!}{\includegraphics[angle=-90]{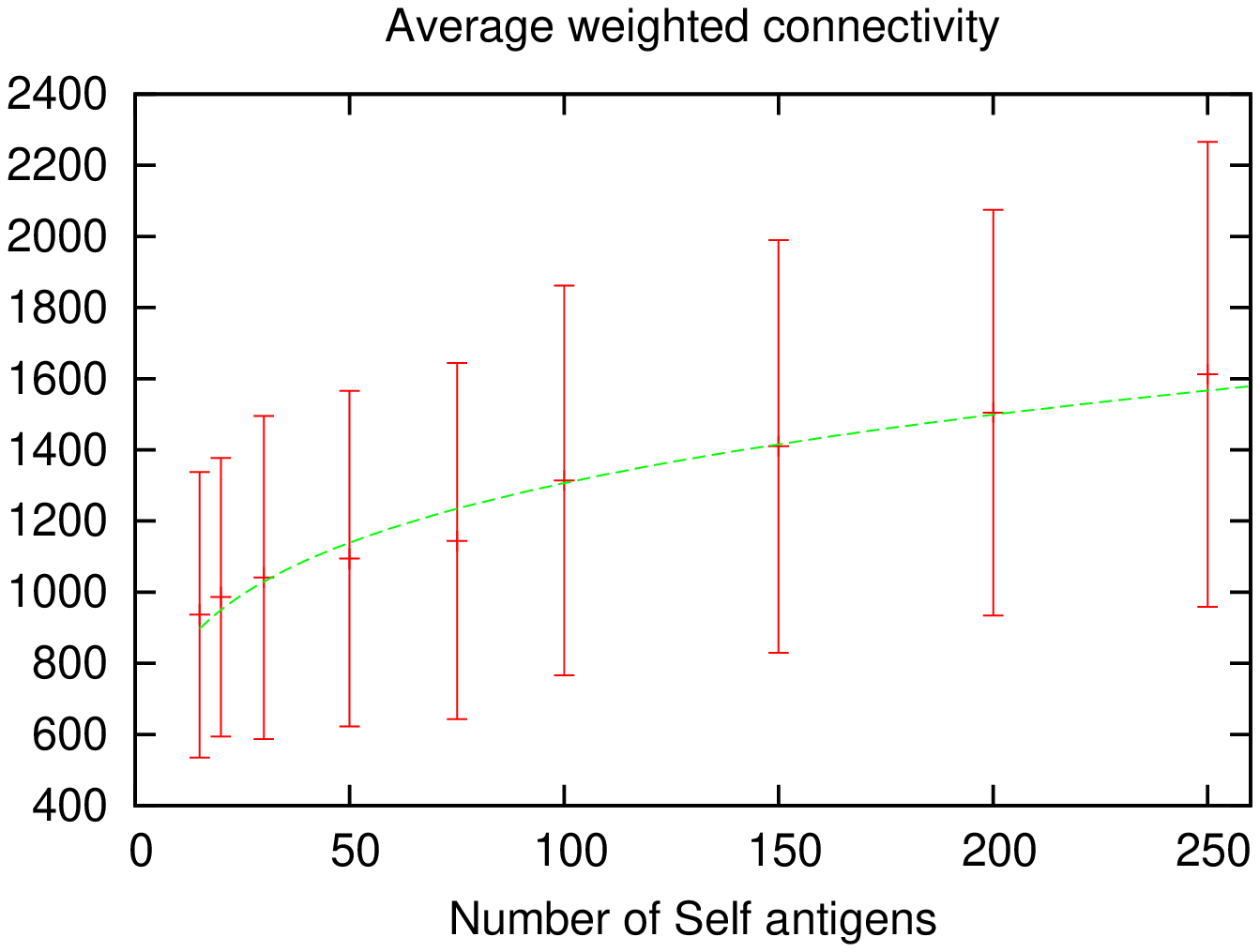}}
\resizebox{0.50\columnwidth}{!}{\includegraphics[angle=-90]{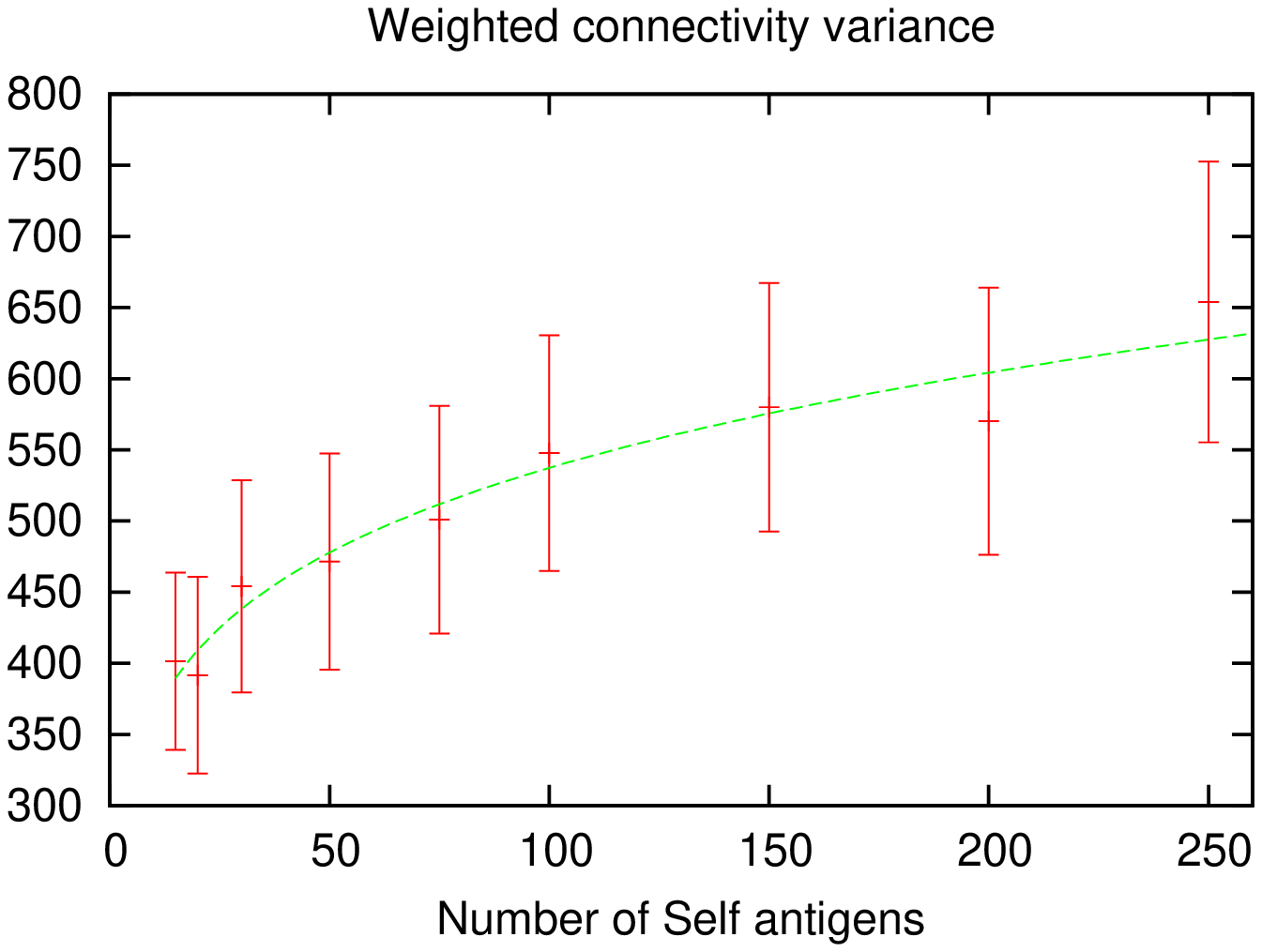}}
\caption{Left: Averaged weighted connectivity for different
repertoires generated increasing the size of the experienced self
$N_S$ at ontogenesis. Right: Standard deviation of the averaged
connectivity plotted at left. We stress that both these quantities
increases with $N_S$, suggesting relations among learning in
ontogenesis  and improved performances in mature behavior. The
power-laws $N_S^{W},N_{S}^{\sigma_{W}}$ obtained by the fits
respectively
 for the connectivity and its variance gave $W=0.19\pm 0.02, \sigma_W=0.17\pm 0.02$.} \label{W}
\end{figure}
Furthermore, as shown in Fig.$(3.1)$ (left),  we notice that
higher the number of self-antigens $N_S$ stored by the system (at
fixed repertoire size $N$), sharper its epitopal matching in
binding antigens as only a small amount of highly affine clones
are available to responde. This can be a cross feature of the
ontogenesis: a direct cause is the generation of holes in the
repertoire, such that, the larger the hole/s the harder the
ability in binding; however, there is another feature ongoing:
 both the average weighed connectivity and its
variance become higher as $N_S$ increases (see Fig. $(3.2)$ left
and right); the whole highlighting a non trivial effect of this
learning process into the mature clonal network: As $N_S$ grows
both these quantities grow\footnote{The similar behavior of the
growth process among average and variance should not surprise as
these networks are random and not too far from the simplest
Erdos-Renyi ones \cite{AB1}, i.e. Poissonian.}. This growth with $N_S$ can be
a very important point because, with respect to a standard random
network with no learning process (i.e. $N_S=0$), this system
displays a larger variance of a (larger) averaged weighted
connectivity: From Varela perspective \cite{a38,a39} this allows a
better response of the mature B-cells against the antigens and  a
stronger anergy for those clones self-directed.
\newline
Coherently with the previous picture of this artificial immune
system, we note that as $N_S$ gets bigger the amount of responding
clones get smaller since, as their average connectivity is
increasing, their responses become weaker, in perfect agreement
with the Varela picture.
\newline
Another interesting observed result is the idiotypic nature of
such a network (not shown): as the responding clones become
activated, they induce activation to the ones with higher
complementary to them (Jerne images of the antigens) and, while
this activation does not propagate extensively trough the system
(only dimers and four loops are observed with our system sizes),
these images actually are found to participate in keeping memory
of the antigen, once it is removed.
\newline
To check abilities in storing memories of the past infections in
this artificial immune network, keeping in mind that trough MC
simulations we only access equilibrium information, we collect
snapshots of the system and confront those representing it
immediately after the antigen infection (when all the responding
clones and their idiotypic counterparts are activated) and the
ones representing the system when the antigen has been removed and
the network stabilized again. As shown in Fig. $(3.3)$ (for two
different examples, i.e. exposures to one (left) or five (right)
 antigens simultaneously), the system maintains (the proper) clones
activated even after the explicit presence of the stimuli has been
removed.
\newline
Of course the resulting antibody (or lymphocyte) concentrations
are lower with respect to the first response as these are thought
of as only memory cells (bridging what in physics is called as
remanent magnetization, an hysteresis effect\footnote{In this
model there is no a-priori difference among plasma and memory
B-cells; the latter are simply though of as remanent
magnetizations, such that, once the antigen is removed -say
$h_k=0$-, its corresponding magnetization $m_k \geq 0$, using the
hysteresis as a generator of dynamical memory trough the
network.}).
\begin{figure}[tb]
%\resizebox{0.50\columnwidth}{!}{\includegraphics{memory1200-1antigen-type2.eps}}
%\resizebox{0.50\columnwidth}{!}{\includegraphics{memory1200-5antigens-type2.eps}}
\resizebox{0.50\columnwidth}{!}{\includegraphics[angle=-90]{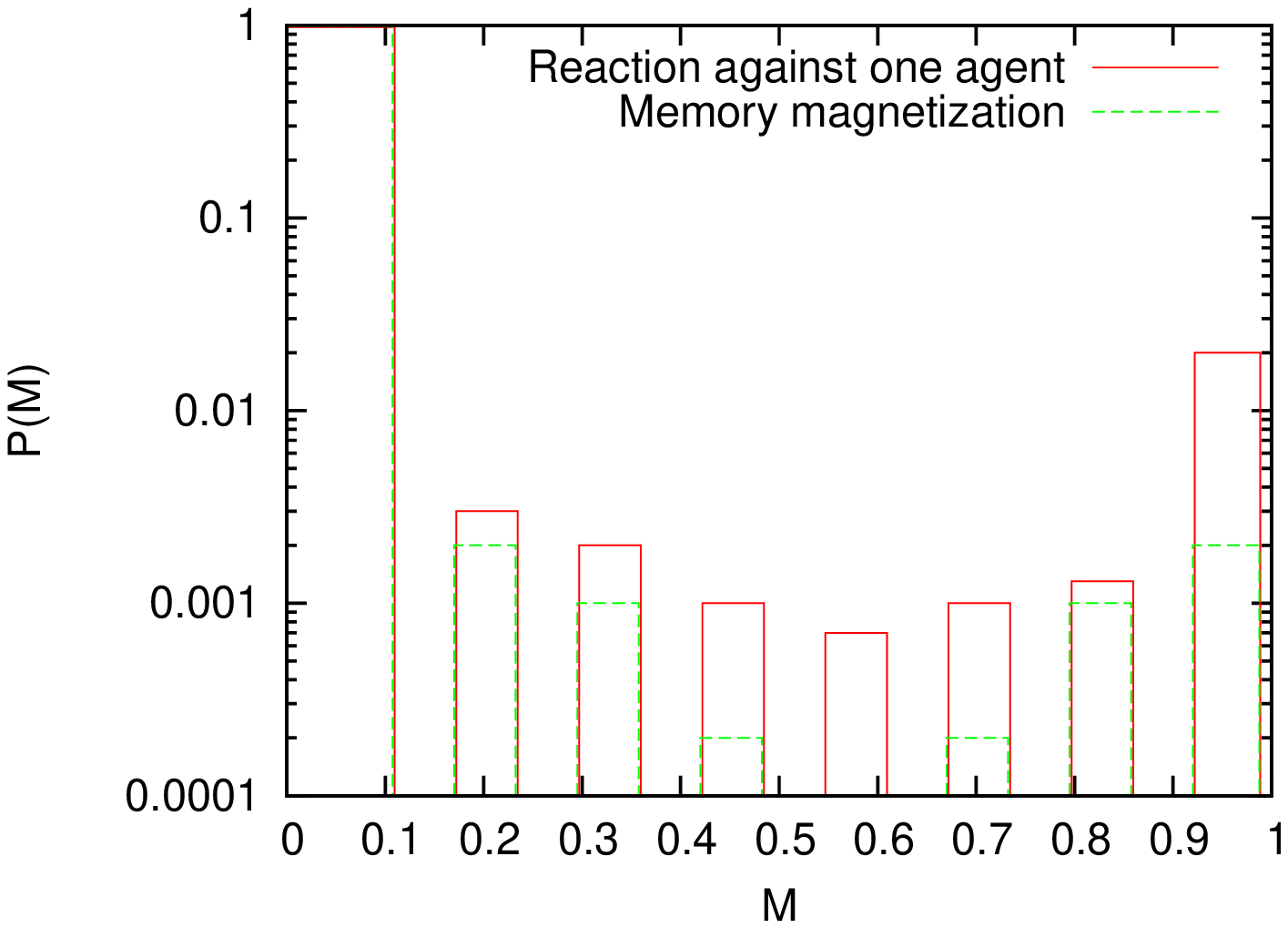}}
\resizebox{0.50\columnwidth}{!}{\includegraphics[angle=-90]{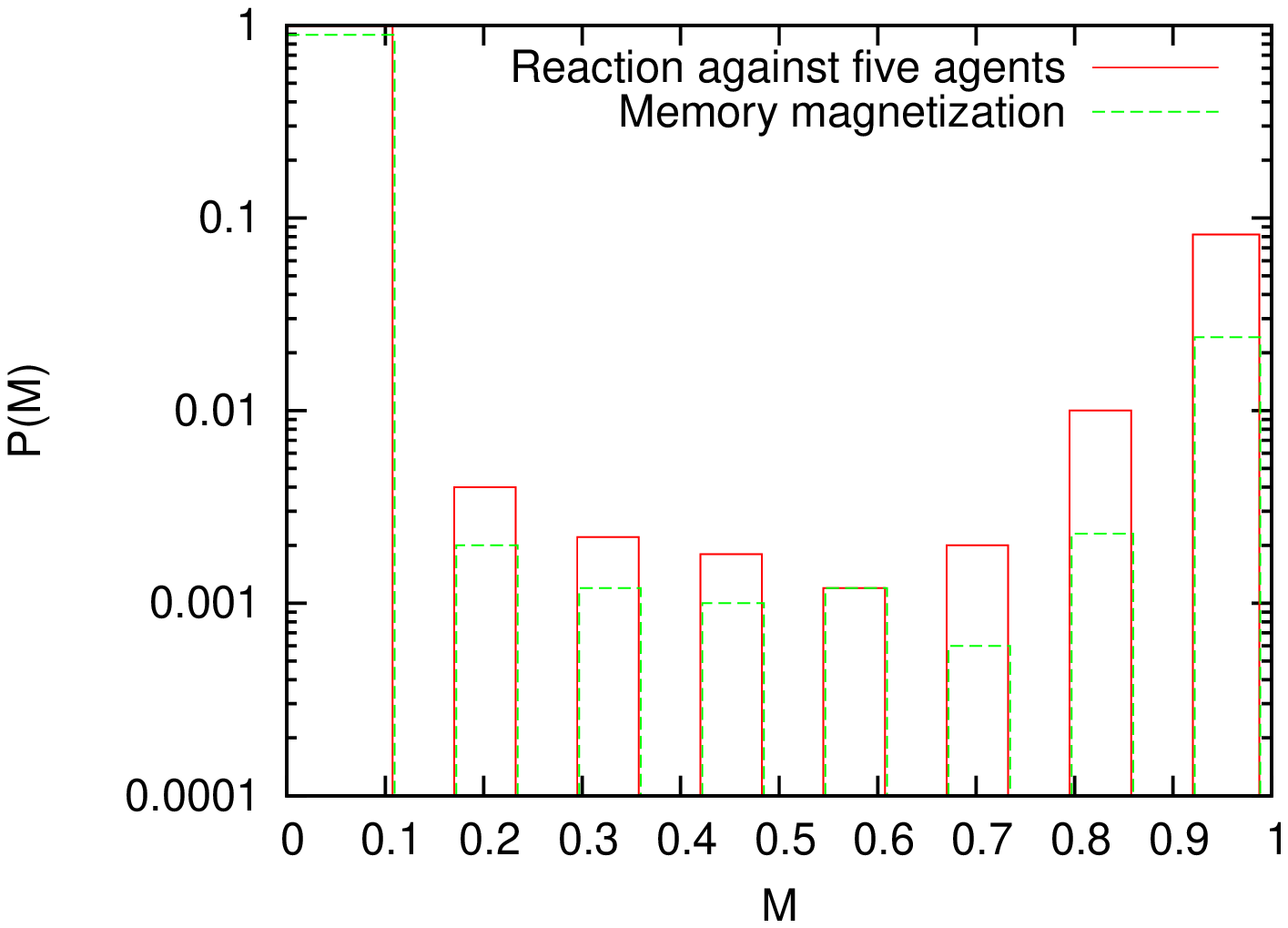}}
\caption{Left: Distribution of the magnetizations during one
antigen attack (red) and after its removal and the successive
network equilibration (green). Right: Distribution of the
magnetization during five contemporary attacks (red) -by different
antigens- and after their removal and the successive network
equilibration (green). The examples show a system made of by
$1257$ clones and $20$ self-antigens.} \label{sette}
\end{figure}
%\begin{figure}[tb]
%\resizebox{0.50\columnwidth}{!}{\includegraphics{memory1200-1antigen-type2.eps}}
%\resizebox{0.50\columnwidth}{!}{\includegraphics{memory1200-5antigens-type2.eps}}
%\caption{blablabla.} \label{sette}
%\end{figure}
%
However, despite the system is extremely able to respond sharply
to the desired infection (or trough a proper best fitting antibody
or trough a linear combination of the "enough matching" ones from
the repertoire, due to hysteresis (which are unavoidable features
as they ensure the dynamical memory), infection after infection,
the system starts behaving plastically and eventually, after a
certain threshold unwanted activations may appear (reflecting
senile autoimmunity) and if this iteration continues, it stops
working at all (no recognition is possible any longer).
\newline
The fraction of the activated clones as a function of the
different antigens continuously experienced is reported in
Fig.$(3.4)$:
\newline
This percolation activation can be easily understood from the
perspective of spin glasses (as well as its aging properties
previously discussed) due to its strong analogy to a diluted
random field model into a magnetic field: since the system works
at low temperature, it undergoes a first order phase transition
for a critical value of the external field \cite{MPV}\cite{silvio2}.
\begin{figure}[tb]
\resizebox{0.50\columnwidth}{!}{\includegraphics[angle=-90]{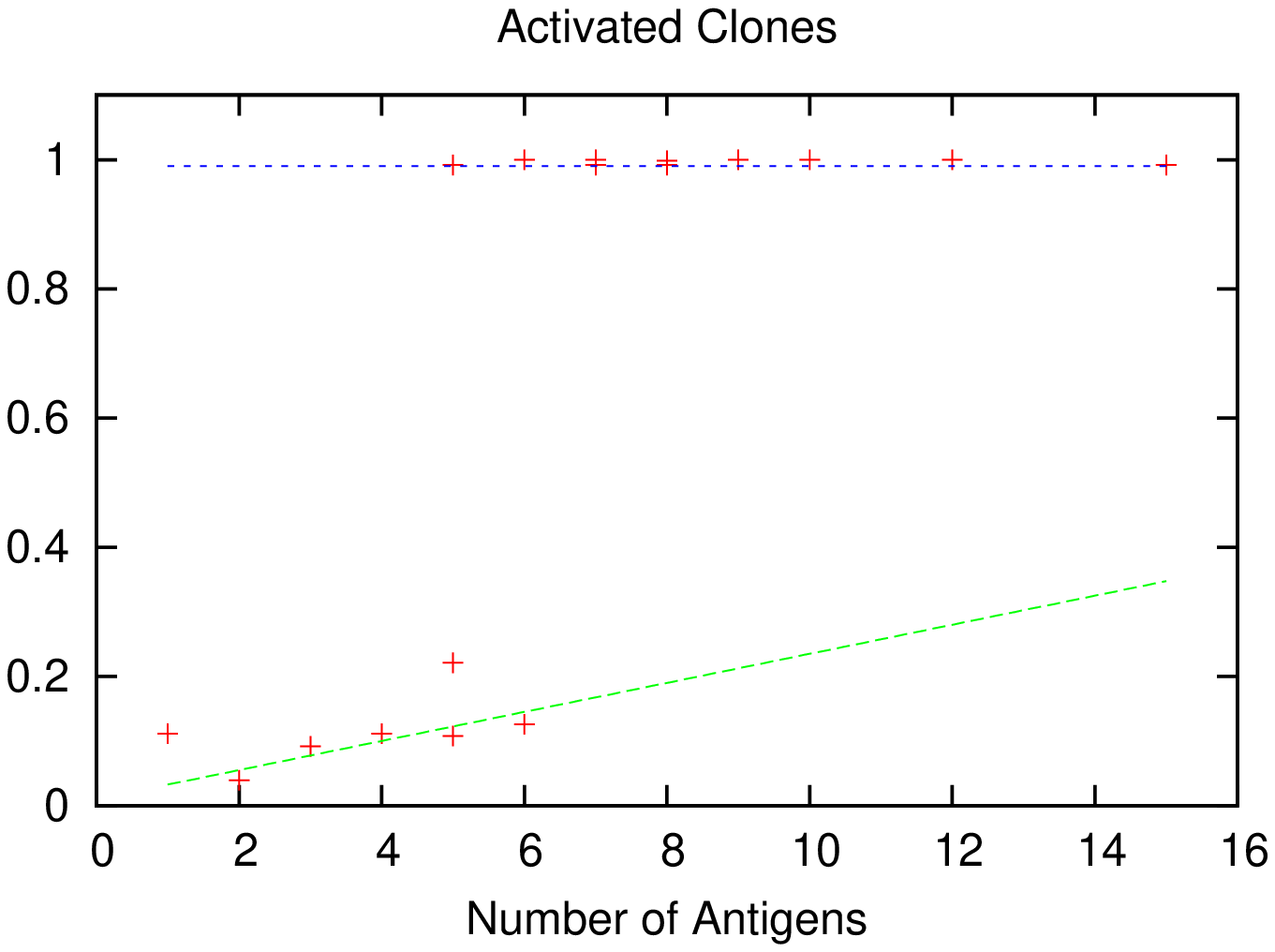}}
\resizebox{0.50\columnwidth}{!}{\includegraphics[angle=-90]{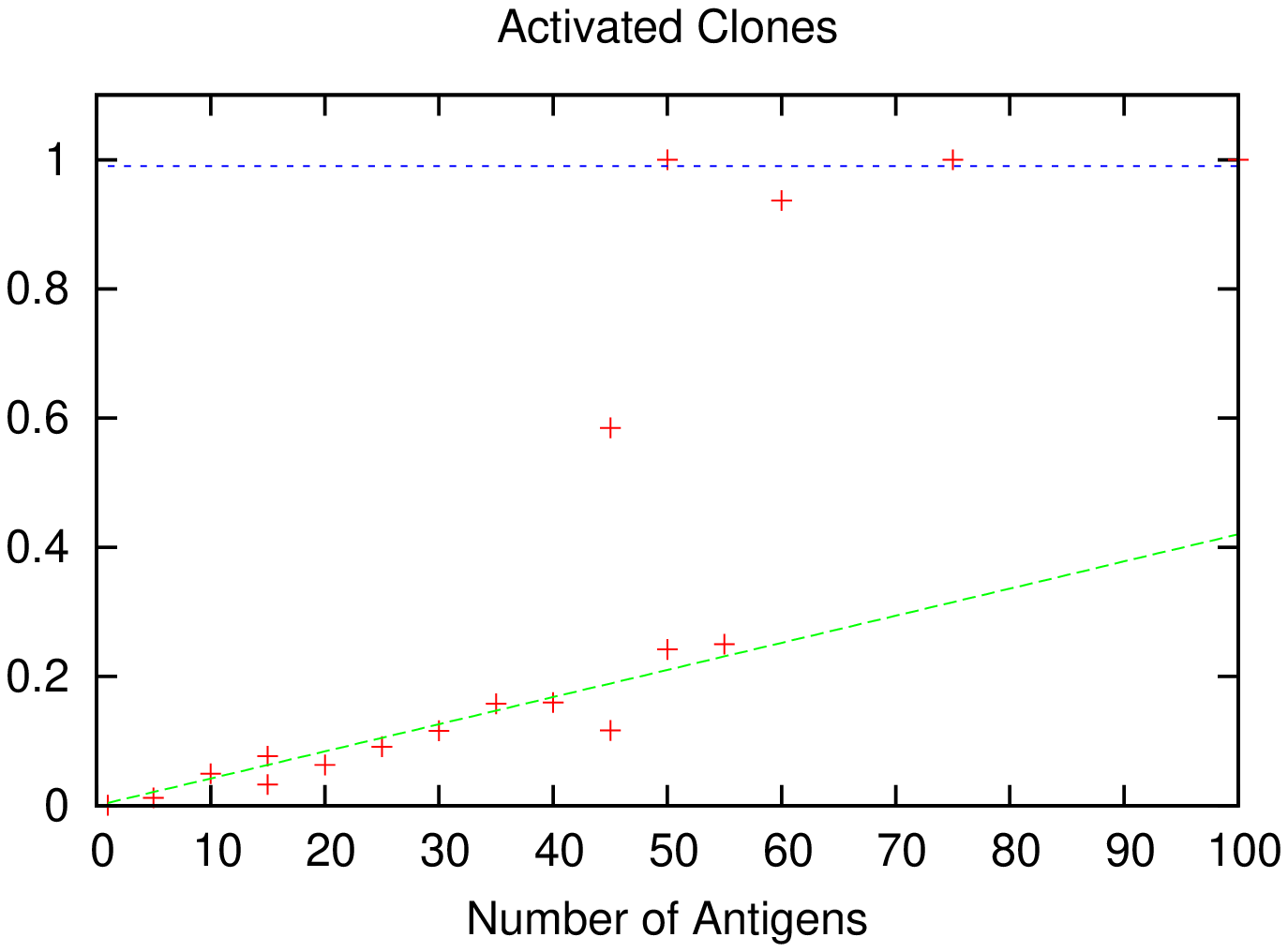}}
\caption{Fraction of the activated clones as a function of the
antigens presented to the system. Left: The system is made of by
$611$ clones and $10$ self-antigens. Right: The system is made of
by $3352$ clones and $50$ self-antigens.} \label{sette}
\end{figure}

\section{Discussion}

In this paper we investigated the ability of memorizing -and its
consequences- for a model of B-cell interactions only. The main
goal was a satisfactory picture by which negative selection
mechanism at ontogenesis may act and operate synergically with the
idiotypic network regulation (for the mature system) in
self/non-self discrimination.
\newline
It is worth noticing that key mechanisms, as helper double
signalling, are fundamental for a complete discrimination process,
however, we investigated this "sub-shell" of the system alone to
highlight features which can result purely by its actions and
which can be more difficult to be revealed when looking at the
 system as a whole. As a result our goal is not meant as an explanation
of the main strand in self/non-self discrimination, but an
investigation of the mechanisms which can participate stemming
from the network perspective.
\newline
In fact, from one side it is now widely accepted the key role of T
helpers in such a regulation: silencing of self-reactive B cells
must be initiated by the binding of self-antigens and because a
B-cell alone cannot distinguish between self and non-self,  the
decision to become anergic must be based on whether secondary
signals are received within a specific time frame: if not, the
cell undergoes BCR desensitization, which ultimately results in
anergy. However, as biological systems, for structural stability,
rarely allow only a single pattern of realization of a macroscopic
behavior, we decided to isolate, in numerical simulations, the
B-cell network and investigate the relations among its ontogenesis
and its mature behavior connected with the problem of
self/non-self discrimination.
\newline
Assuming a random repertoire for the self-antigen, we showed that,
at first, the system is able to learn these antigens and to avoid
attacking them even at successive equilibrium, then, we showed
that this learning mechanism, on the experienced self, increases
the average weighted connectivity of the resulting network of
interacting lymphocytes as well as its variance: this is an
important bridge among ontogenesis and mature repertoire as, from
initial clonal deletion, the system can manage more reactive
antibodies against the pathogens (low connected clones) and more
anergic self-directed ones (of course within the Varela and
Coutinho viewpoint), confirming both the possibility and the
utility of the two mechanisms.
\newline
Future development should include T-helper interactions (which may
spread the phenomenon on several time-scales due to the intrinsic
 three-body interactions on
 diluted network \cite{silvio}) as well as a systematic
 exploration of the relation among the amount of stored self-antigens in ontogenesis with respect to stability of the mature response against the number of encountered pathogens, so to understand the stability region of this system, where it works as a pattern reconstructor,
 with respect to its breakdown (so to try a statistical mechanics approach to memory saturation in the immune networks).

\section*{Acknowledgements}

The authors are pleased to thank  Elena Agliari, Francesco Guerra,
Pietro Lio' and Luca Peliti for useful discussions.
\newline
Part of the research belongs to the strategy of exploration funded
by the FIRB project $RBFR08EKEV$ which is acknowledged.

\end{document}